\documentclass[prl,aps,nofootinbib,notitlepage,twocolumn,superscriptaddress]{revtex4-2}

\usepackage{amsthm}
\usepackage{amsmath,bm}
\usepackage{amssymb}
\usepackage{amsfonts}
\usepackage{graphicx}

\usepackage{txfonts}

\usepackage[colorlinks=true,linkcolor=blue,citecolor=blue,urlcolor=blue]{hyperref}

\begin{document}

\title{Hierarchies of Frequentist Bounds for Quantum Metrology: From Cram\'er-Rao to Barankin}
\author{Manuel Gessner}
\email{manuel.gessner@uv.es}
\affiliation{Departament de F\'isica Te\`orica, IFIC, Universitat de Val\`encia, CSIC, C/ Dr. Moliner 50, 46100 Burjassot (Val\`encia), Spain}
\author{Augusto Smerzi}
\email{augusto.smerzi@ino.it}
\affiliation{QSTAR, INO-CNR and LENS, Largo Enrico Fermi 2, 50125 Firenze, Italy}
\date{\today}

\begin{abstract}
    We derive lower bounds on the variance of estimators in quantum metrology by choosing test observables that define constraints on the unbiasedness of the estimator. The quantum bounds are obtained by analytical optimization over all possible quantum measurements and estimators that satisfy the given constraints. We obtain hierarchies of increasingly tight bounds that include the quantum Cram\'er-Rao bound at the lowest order. In the opposite limit, the quantum Barankin bound is the variance of the locally best unbiased estimator in quantum metrology. Our results reveal generalizations of the quantum Fisher information that are able to avoid regularity conditions and identify threshold behavior in quantum measurements with mixed states, caused by finite data.
\end{abstract}

\maketitle
\textit{Introduction.---}Quantum metrology provides the theory foundation for identifying precision limits and developing quantum-enhanced strategies in quantum measurements~~\cite{HelstromBOOK,HolevoBOOK,ParisINTJQI2009,GiovannettiNATPHOT2011,TothJPA2014,PezzeRMP2018}. Today, quantum sensing and metrology represent some of the most advanced quantum technologies with applications ranging from gravitational wave detection~\cite{TsePRL2019} to atomic clocks and interferometers~\cite{PezzeRMP2018}. The cornerstone of frequentist quantum metrology, the quantum Cram\'er-Rao bound (QCRB), identifies a lower bound on the variance of an estimator for a parameter encoded in a quantum state~\cite{BraunsteinPRL1994}. The inverse of the QCRB is a measure for the quantum state's sensitivity under small variations of a parameter, known as the quantum Fisher information (QFI). Nowadays, the QFI is used, far beyond its original purpose in quantum metrology, as an extremely powerful and versatile tool in quantum information theory. For instance, it efficiently detects multipartite entanglement~\cite{PezzeSmerziReview,RenPRL2021}, Einstein-Podolsky-Rosen steering~\cite{YadinNatCommun2021}, quantum phase transitions~\cite{ZanardiPRL2007,HaukeNatPhys2016,PezzePRL2017}, non-Markovian open-system evolutions~\cite{NonMarkov}, and allows to sharpen uncertainty relations~\cite{BCMAnnalsPhysics1996,Uncertainty}, quantum speed limits~\cite{TaddeiPRL2013}, and the quantum Zeno effect~\cite{SmerziPRL2012}.

The QCRB can be understood as the natural quantization of the classical CRB by analytical optimization over all possible quantum measurements~\cite{BraunsteinPRL1994}. As such, the QCRB inherits many properties from its classical counterpart, including its shortcomings. For instance, the CRB is undefined for estimation problems that do not satisfy certain regularity conditions. Furthermore, saturation of the CRB typically only occurs in the asymptotic limit of many repeated measurements, i.e., when the signal-to-noise ratio is large. Consequently, the QCRB provides only very limited information about the achievable precision in few-shot scenarios. To realistically identify limits on the variance of unbiased estimators, tighter bounds are required that generalize the QCRB for low signal-to-noise ratio. In quantum metrology, the unattainability of the QCRB has been pointed out using Bayesian approaches that involve some form of prior information~\cite{TsangPRL2012,LuTsangQST2016,RubioDunningham}. This so-called threshold behavior has been studied in classical information theory within the frequentist paradigm, to which the CRB belongs, through comparison with tighter bounds~\cite{McAulaySeidman1969,McAulayHofstetter1971,Knockaert1997,Renaux2007,Chaumette2008}.

Generalizations of the CRB appeared already in its contemporary literature, most notably the families of bounds by Bhattacharyya~\cite{Bhattacharyya1946} and Barankin~\cite{Barankin1949}. Both families consist of hierarchies of bounds that impose increasingly demanding conditions on the unbiasedness of an estimator. The Bhattacharyya bounds compare the estimator to the identity function in a neighborhood of the unknown parameter's true value by means of Taylor-like expansion involving higher-order derivatives. The Barankin bounds impose unbiasedness of the estimator at an increasing number of $n$ test points within the range of possible values for the unknown parameter. In the limit $n\to\infty$, the Barankin bounds identify the variance of the locally best unbiased  estimator~\cite{Barankin1949}. By avoiding derivatives altogether, the Barankin bounds circumvent the regularity conditions of the CRB and, more generally, the Bhattacharyya bounds. The Abel bounds~\cite{Abel1993} combine the ideas of Bhattacharyya and Barankin and involve a combination of test points and higher-order derivatives. In all cases, these approaches produce hierarchies of bounds that are tighter than the CRB, which in turn is recovered at the respective lowest order.

In this Letter, we derive a hierarchy of increasingly tight bounds for quantum parameter estimation that include the QCRB as a special case at the lowest order. Like the QCRB, all generalized quantum bounds are obtained from classical frequentist bounds by an analytical optimization over all possible quantum measurements. Each bound is determined by a choice of test observables that reflect generalized unbiasedness conditions on the estimator. In the limit of pure states, the hierarchy collapses and all bounds reproduce the QCRB. As relevant applications, we show how our approach naturally quantizes the bounds by Bhattacharyya~\cite{Bhattacharyya1946}, Barankin~\cite{Barankin1949}, Hammersley~\cite{Hammersley1950}, Chapman and Robbins~\cite{Chapman1951}, and Abel~\cite{Abel1993}. We provide explicit expressions for the bounds and the quantum information functions, i.e., generalizations of the QFI based on the Bures metric, that are defined by their inverses. We illustrate our results with an analysis of threshold behavior in quantum phase estimation with a qubit.

\textit{Hierarchy of quantum bounds.---}We consider an estimator $\theta_{\mathrm{est}}(x)$ for $\theta$ that is a function of the obtained measurement results $x$. For a fixed measurement setting, described by the positive operator-valued measure (POVM) $E_x\geq 0$, with $\sum_xE_x=\mathbb{I}$, the probability to obtain the result $x$ when the parameter of interest takes on the value $\theta$ is given by $p(x|\theta)=\mathrm{Tr}\{E_x\rho(\theta)\}$, where $\rho(\theta)$ is the quantum state of the system. In the following, we describe the state of the measurement apparatus with a generic (single-copy) density matrix $\rho(\theta)$. The $m$-shot scenario is included in this description by replacing $\rho(\theta)$ with $\rho(\theta)^{\otimes m}$.

The quantity of central interest is the variance of the estimator,
\begin{align}\label{eq:varest}
    (\Delta\theta_{\mathrm{est}})^2=\sum_{x} p(x|\theta)\left(\theta_{\mathrm{est}}(x)-\langle\theta_{\mathrm{est}}\rangle_{\theta}\right)^2,
\end{align}
with the average value $\langle\theta_{\mathrm{est}}\rangle_{\theta}=\sum_{x}p(x|\theta)\theta_{\mathrm{est}}(x)$. We are interested in the locally best estimators, i.e., those that minimize the variance when the unknown parameter takes on the value $\theta$. We consider estimators that are unbiased over the range $\Theta\subset\mathbb{R}$ of possible parameter values, i.e., $\langle\theta_{\mathrm{est}}\rangle_{\theta_0}=\theta_0$ for all $\theta_0\in\Theta$. The starting point for our derivation of lower bounds on~(\ref{eq:varest}) is the choice of a family of test observables $\bm{G}=(G_1,\dots,G_n)^{\top}$. Each $G_k$ is a Hermitian operator on the same Hilbert space as $\rho(\theta)$. Typically, $G_k$ will be a linear function of $\rho(\theta_k)$ where $\theta_k\in\Theta$. In the following, we derive generalized quantum bounds for any choice of the $G_k$, and further below, we show how suitable choices for the $G_k$ lead to a hierarchy of bounds that generalize the QCRB. The number $n$ of different test observables $G_k$ determines the order of the bound. Each $G_k$ allows us to impose an unbiasedness constraint $\lambda_k$ on the estimator $\theta_{\mathrm{est}}$.

In combination with a measurement of the POVM element $E_{x}$, the test observables give rise to the classical functions $g_k(x)=\mathrm{Tr}\{E_{x}G_k\}$. We first derive classical bounds that hold for any choice of the test functions $\bm{g}(x)=(g_1(x),\dots,g_n(x))^{\top}$ and in a second step optimize these bounds over all choices of POVMs. The bounds are based on linear combinations of the test functions of the form $\bm{a}^{\top}\bm{g}(x)=\sum_{k=1}^na_kg_k(x)$, where $\bm{a}=(a_1,\dots,a_n)^{\top}\in\mathbb{R}^n$ is a vector of coefficients. We find that $(\Delta\theta_{\mathrm{est}})^2\geq(\Delta\theta_{\mathrm{est}})_C^2$, with~\cite{Supp}
\begin{align}\label{eq:cBound}
    (\Delta\theta_{\mathrm{est}})_C^2 = \sup_{\bm{a}} \frac{\left(\bm{a}^{\top}\bm{\lambda}\right)^2}{\bm{a}^{\top}C\bm{a}}=\bm{\lambda}^{\top}C^{-1}\bm{\lambda},
\end{align}
where the second equality holds when $C$ is invertible. We have introduced the bias conditions $\bm{\lambda}=(\lambda_1,\dots,\lambda_n)^{\top}$ with
\begin{align}\label{eq:lambdak}
    \lambda_k=\sum_{x\in X_+}g_k(x)(\theta_{\mathrm{est}}(x)-\langle\theta_{\mathrm{est}}\rangle_{\theta})
\end{align}
and $C$ is a real, symmetric $n\times n$ classical information matrix with elements
\begin{align}
    C_{kl}=\sum_{x\in X_+}\frac{g_k(x)g_l(x)}{p(x|\theta)}.
\end{align}
The sums are restricted to those events $x\in X_+$ that occur with nonzero probability $p(x|\theta)>0$.

To obtain the corresponding quantum bounds, we now minimize the right-hand side of Eq.~(\ref{eq:cBound}) over all POVMs $\{E_x\}$. We assume that the $\lambda_k$ are fixed properties of the estimator that characterize the bias. Thus, each test observable $G_k$ imposes another constraint~(\ref{eq:lambdak}) on the unbiasedness of $\theta_{\mathrm{est}}$. The analytical optimization over all POVMs yields $(\Delta\theta_{\mathrm{est}})^2\geq~(\Delta\theta_{\mathrm{est}})_C^2\geq (\Delta\theta_{\mathrm{est}})_Q^2$ with~\cite{Supp}
\begin{align}\label{eq:qBound}
    (\Delta\theta_{\mathrm{est}})_Q^2=\min_{\{E_x\}}(\Delta\theta_{\mathrm{est}})_C^2=\sup_{\bm{a}}\frac{\left(\bm{a}^{\top}\bm{\lambda}\right)^2}{\bm{a}^{\top}Q\bm{a}}=\bm{\lambda}^{\top}Q^{-1}\bm{\lambda},
\end{align}
where, again, the final equality in Eq.~(\ref{eq:qBound}) holds for $Q$ invertible, and we obtain the real, symmetric $n\times n$ quantum information matrix $Q$ with elements
\begin{align}\label{eq:qMatrix}
    Q_{kl}=\mathrm{Tr}\{G_k\Omega_{\rho(\theta)}(G_l)\}.
\end{align}
The operator $\Omega_A(X)$ is defined by the property
\begin{align}\label{eq:symdivision}
    X=\frac{A\Omega_{A}(X)+\Omega_{A}(X)A}{2}.
\end{align}
Intuitively, $\Omega_A(X)$ describes a symmetric "division" of $X$ by the operator $A$. Furthermore, the bound~(\ref{eq:qBound}) is saturated by an optimal projective measurement whose elements $E_x$ are the projectors onto the eigenstates of $\Omega_{\rho(\theta)}(\bm{a}^{\top}\bm{G})$. The optimal coefficient vector $\bm{a}$ achieves the supremum in Eq.~(\ref{eq:qBound}) and is given by $\bm{a}=\alpha Q^{-1}\bm{\lambda}$ when $Q^{-1}$ exists, with $\alpha\in\mathbb{R}$ a normalization constant.

The bound~(\ref{eq:qBound}) and definition~(\ref{eq:symdivision}) apply to test observables that satisfy $\Pi_{\perp}G_k \Pi_{\perp}=0$ for all $k$, where $\Pi_{\perp}$ is the projector onto the eigenspace of $\rho(\theta)$ with eigenvalue $0$. This condition can always be satisfied by adding $\epsilon\mathbb{I}$ to $\rho(\theta)$ with $\epsilon\ll 1$. A discussion of technical details and saturation conditions is provided in Ref.~\cite{Supp}.

To illustrate our approach, consider the simple scenario of a single test observable ($n=1$) chosen as $G_1=\frac{\partial \rho(\theta)}{\partial \theta}$. It is easy to see that the bound~(\ref{eq:cBound}) yields the classical CRB with bias condition $\lambda_1=\frac{\partial}{\partial \theta}\langle\theta_{\mathrm{est}}\rangle_{\theta}$ and the Fisher information $C$. The associated quantum bound~(\ref{eq:qBound}) is the well-known QCRB, i.e., $Q$ yields the quantum Fisher information $F_Q[\rho(\theta)]$. More general choices of the test observables reveal a rich variety of quantum bounds that provide more realistic estimates of the smallest achievable variance of unbiased estimators than the QCRB. The result~(\ref{eq:qBound}) also includes special cases that are discussed in Refs.~\cite{Tsuda2005}.

\textit{Quantum Barankin bounds.---}Consider the family of test observables $G_k=\rho(\theta_k)$, representing the density matrix at different values of the parameter $\theta$ within the range of possible parameters, called test points $\theta_k\in\Theta$. We are interested in estimators that are unbiased at the true value $\theta$ and at each of these test points, $\langle\theta_{\mathrm{est}}\rangle_{\theta_k}=\theta_k$, which in~(\ref{eq:lambdak}) yield $\lambda_k=\theta_k-\theta$. 

Classical bounds for estimators that are unbiased at $n$ arbitrary test points within the range $\Theta$ were first derived by Barankin in 1949~\cite{Barankin1949}. The Barankin bound (BaB) of order $n$ is recovered from Eq.~(\ref{eq:cBound}) for the choice $\bm{G}=(\rho(\theta_1),\dots,\rho(\theta_n))^{\top}$, after a maximization over the choice of test points $\lambda_k$, and reads $(\Delta\theta_{\mathrm{est}})_{C_{\mathrm{Ba}}}^2 = \sup_{\bm{\lambda}}\bm{\lambda}^{\top}C_{\mathrm{Ba}}^{-1}\bm{\lambda}$, with the Barankin information matrix
\begin{align}
    (C_{\mathrm{Ba}})_{kl}=\sum_{x\in X_+}p(x|\theta)L(x|\theta+\lambda_k,\theta)L(x|\theta+\lambda_l,\theta),
\end{align}
and $L(x|\theta+\lambda_k,\theta)=\frac{p(x|\theta+\lambda_k)}{p(x|\theta)}$ is the likelihood ratio. Barankin further showed that an additional optimization of all BaBs over the number $n$ of test points (or equivalently $n\to\infty$) yields the variance of the locally best unbiased estimator, which is unique~\cite{Barankin1949}. The Barankin bound is typically hard to determine but has the advantage of avoiding the regularity conditions of the CRB. Efficient approximations of the Barankin bound based on small $n$ have proven to be useful to highlight the limitations of the CRB at small signal-to-noise ratio~\cite{McAulaySeidman1969,McAulayHofstetter1971,Knockaert1997,Chaumette2008}.

We identify the quantum Barankin bounds (QBaBs) by optimizing over all possible POVMs, which according to Eq.~(\ref{eq:qBound}) yields $\min_{\{E_x\}}(\Delta\theta_{\mathrm{est}})_{C_{\mathrm{Ba}}}^2=\sup_{\bm{\lambda}}\bm{\lambda}^{\top}Q_{\mathrm{Ba}}^{-1}\bm{\lambda}$ with the quantum Barankin information (QBaI) matrix
\begin{align}\label{eq:qBaMatrix}
    (Q_{\mathrm{Ba}})_{kl}=\mathrm{Tr}\{\rho(\theta+\lambda_k)\Omega_{\rho(\theta)}(\rho(\theta+\lambda_l))\}.
\end{align}
The QBaBs can be improved at no additional computational cost by adding $\rho(\theta_0)=\rho(\theta)$ to the set $\bm{G}$~\cite{McAulaySeidman1969}, with the corresponding bias constraint $\lambda_0=0$, leading to~\cite{Supp}
\begin{align}\label{eq:qBarankin}
    (\Delta\theta_{\mathrm{est}})_{Q_{\mathrm{Ba}}}^2=\sup_{\bm{\lambda}}\bm{\lambda}^{\top}(Q_{\mathrm{Ba}}-\bm{e}\bm{e}^{\top})^{-1}\bm{\lambda},
\end{align}
where $\bm{e}=(1,\dots,1)^{\top}\in\mathbb{R}^n$.

At its lowest order, $n=1$, the classical BaB reduces to the Hammersley-Chapman-Robbins bound (HCRB)~\cite{Hammersley1950,Chapman1951}, which reads
\begin{align}
    (\Delta\theta_{\mathrm{est}})^2_{C_\mathrm{HCRB}}=\sup_{\lambda}\frac{\lambda^2}{\chi^2[p(\cdot|\theta+\lambda),p(\cdot|\theta)]}.
\end{align}
Here, $\chi^2[p(\cdot|\theta+\lambda),p(\cdot|\theta)]=\sum_{x\in X_+}\frac{p(x|\theta+\lambda)^2}{p(x|\theta)}-1$ is the $\chi^2$-divergence of $p(\cdot|\theta+\lambda)$ with respect to $p(\cdot|\theta)$. The quantum Hammersley-Chapman-Robbins bound (QHCRB), $(\Delta\theta_{\mathrm{est}})^2_{Q_{\mathrm{HCRB}}}=\min_{\{E_{x}\}}(\Delta\theta_{\mathrm{est}})^2_{C_\mathrm{HCRB}}$, i.e., the QBaB~(\ref{eq:qBarankin}) at $n=1$ reads
\begin{align}\label{eq:qHCRB}
    (\Delta\theta_{\mathrm{est}})^2_{Q_{\mathrm{HCRB}}}=\sup_{\lambda}\frac{\lambda^2}{\mathrm{Tr}\{\rho(\theta+\lambda)\Omega_{\rho(\theta)}(\rho(\theta+\lambda))\}-1}.
\end{align}
The denominator indeed coincides with the quantum $\chi^2$-divergence $\chi_Q^2[\rho(\theta+\lambda),\rho(\theta)]=\max_{\{E_{x}\}}\chi^2[p(\cdot|\theta+\lambda),p(\cdot|\theta)]=\mathrm{Tr}\{\rho(\theta+\lambda)\Omega_{\rho(\theta)}(\rho(\theta+\lambda))\}-1$~\cite{TemmeJMP2015}.

Already at this lowest order, the QBaBs generalize the QCRB: If $\rho(\theta)$ is differentiable, we recover the QCRB from the right-hand side of Eq.~(\ref{eq:qHCRB}) by replacing the supremum with the limit $\lambda\to 0$~\cite{TemmeJMP2015,Supp}. Generally, however, the QHCRB is tighter, $(\Delta\theta_{\mathrm{est}})^2_{Q_{\mathrm{HCRB}}}\geq (\Delta\theta_{\mathrm{est}})^2_{Q_{\mathrm{CRB}}}=F_Q[\rho(\theta)]^{-1}$, and applicable to a wider range of problems.

\textit{Quantum Bhattacharyya bounds.---}Let us consider the test observables $G_k=\frac{\partial^k\rho(\theta)}{\partial\theta^k}$, assuming that these derivatives exist. The bias conditions~(\ref{eq:lambdak}) here identify $\lambda_k=\frac{\partial^k}{\partial\theta^k}\langle\theta_{\mathrm{est}}\rangle_{\theta}$ for $k>0$ as fixed properties of the estimator. Unbiased estimators satisfy $\lambda_1=1$ and $\lambda_k=0$ for all $k>1$.

Choosing the family of test functions $\bm{G}=(\frac{\partial\rho(\theta)}{\partial\theta},\dots,\frac{\partial^n\rho(\theta)}{\partial\theta^n})^{\top}$ with corresponding bias conditions~(\ref{eq:lambdak}) $\lambda=(1,0,\dots,0)^{\top}$ leads in Eq.~(\ref{eq:cBound}) to the Bhattacharyya bound (BhB) of order $n$~\cite{Bhattacharyya1946}, $(\Delta\theta_{\mathrm{est}})_{C_{\mathrm{Bh}}}^2 = \bm{\lambda}^{\top}C_{\mathrm{Bh}}^{-1}\bm{\lambda}$, with the Bhattacharyya information matrix
\begin{align}
    (C_{\mathrm{Bh}})_{kl}=\sum_{x\in X_+}\frac{1}{p(x|\theta)}\left(\frac{\partial^kp(x|\theta)}{\partial\theta^k}\right)\left(\frac{\partial^lp(x|\theta)}{\partial\theta^l}\right).
\end{align}
Conceptually, the BhBs account for the unbiasedness of $\theta_{\mathrm{est}}$ in an increasingly large region around the true value $\theta$ by adding constraints on the higher-order terms of a Taylor expansion.

The quantum Bhattacharyya bound (QBhB) $(\Delta\theta_{\mathrm{est}})_{Q_{\mathrm{Bh}}}^2=\min_{\{E_x\}}(\Delta\theta_{\mathrm{est}})_{C_{\mathrm{Bh}}}^2=\bm{\lambda}^{\top}Q_{\mathrm{Bh}}^{-1}\bm{\lambda}$ follow from Eq.~(\ref{eq:qBound}) with the quantum Bhattacharyya information (QBhI) matrix
\begin{align}\label{eq:qBhMatrix}
    (Q_{\mathrm{Bh}})_{kl}=\mathrm{Tr}\{\frac{\partial^k\rho(\theta)}{\partial\theta^k}\Omega_{\rho(\theta)}\left(\frac{\partial^l\rho(\theta)}{\partial\theta^l}\right)\}.
\end{align}
It is easy to see that at $n=1$, the QBhB coincides with the QCRB, while for higher orders, it is generally tighter. Compared to the QBaB, the QBhB require stronger regularity conditions, since all higher-order derivatives must exist. In contrast, the QBhBs avoid the optimization over the parameters $\bm{\lambda}$ of the QBaB, which can be computationally expensive. Since higher-order derivatives can be recovered from the differences of infinitesimally separated test points, the QBaBs include the QBhBs as special cases~\cite{Barankin1949}.

\textit{Quantum Abel bounds.---}The bounds of Barankin and Bhattacharyya can be combined into hybrid bounds by considering a combination of test observables that contain both $\rho(\theta+\lambda_k)$ and $\frac{\partial^l\rho(\theta)}{\partial\theta^l}$. Such bounds were first discussed by Abel in 1993~\cite{Abel1993}. We obtain the classical Abel bounds of order $(r,s)$, $(\Delta\theta_{\mathrm{est}})_{C_{\mathrm{A}}}^2$, as well as the corresponding quantum Abel bounds (QABs) by considering the family of $r+s$ test observables $\bm{G}=(\rho(\theta+\lambda_1),\dots,\rho(\theta+\lambda_r),\frac{\partial\rho(\theta)}{\partial\theta},\dots,\frac{\partial^s\rho(\theta)}{\partial\theta^s})^{\top}$ in Eqs.~(\ref{eq:cBound}) and~(\ref{eq:qBound}), respectively. The QABs read~\cite{Supp}
\begin{align}\label{eq:QAB}
    (\Delta\theta_{\mathrm{est}})_{{Q}_{\mathrm{A}}}^2=\min_{\{E_x\}}(\Delta\theta_{\mathrm{est}})_{C_{\mathrm{A}}}^2=\sup_{\lambda_1,\dots,\lambda_r}\bm{\lambda}^{\top}(Q_{\mathrm{A}}-\bm{f}\bm{f}^{\top})^{-1}\bm{\lambda},
\end{align}
where $\bm{\lambda}=(\lambda_1,\dots,\lambda_r,1,0,\dots,0)\in\mathbb{R}^{r+s}$ combines the unbiasedness conditions of the Barankin and Bhattacharyya bounds and $\bm{f}=\bm{e}\oplus\bm{0}\in\mathbb{R}^{r+s}$, where $\bm{0}\in\mathbb{R}^s$ is the zero vector. The $(r+s)\times(r+s)$ quantum Abel information (QAI) matrix
\begin{align}\label{eq:QAImatrix}
    Q_{\mathrm{A}}=\begin{pmatrix}Q_{\mathrm{Ba}} & Q_{\mathrm{H}}\\Q^{\top}_{\mathrm{H}}& Q_{\mathrm{Bh}} \end{pmatrix}
\end{align}
contains the $r\times r$ QBaI matrix $Q_{\mathrm{Ba}}$~(\ref{eq:qBaMatrix}) and the $s\times s$ QBhI matrix $Q_{\mathrm{Bh}}$~(\ref{eq:qBhMatrix}), as well as the $r\times s$ hybrid matrix
\begin{align}
    (Q_{\mathrm{H}})_{kl}=\mathrm{Tr}\{\rho(\theta+\lambda_k)\Omega_{\rho(\theta)}\left(\frac{\partial^l\rho(\theta)}{\partial\theta^l}\right)\}.
\end{align}
Even more general bounds can be obtained by including higher-order derivatives also at each of the test points $\theta+\lambda_k$, see, e.g., \cite{Chaumette2008}.

\begin{figure}[tb]
    \centering
    \includegraphics[width=.48\textwidth]{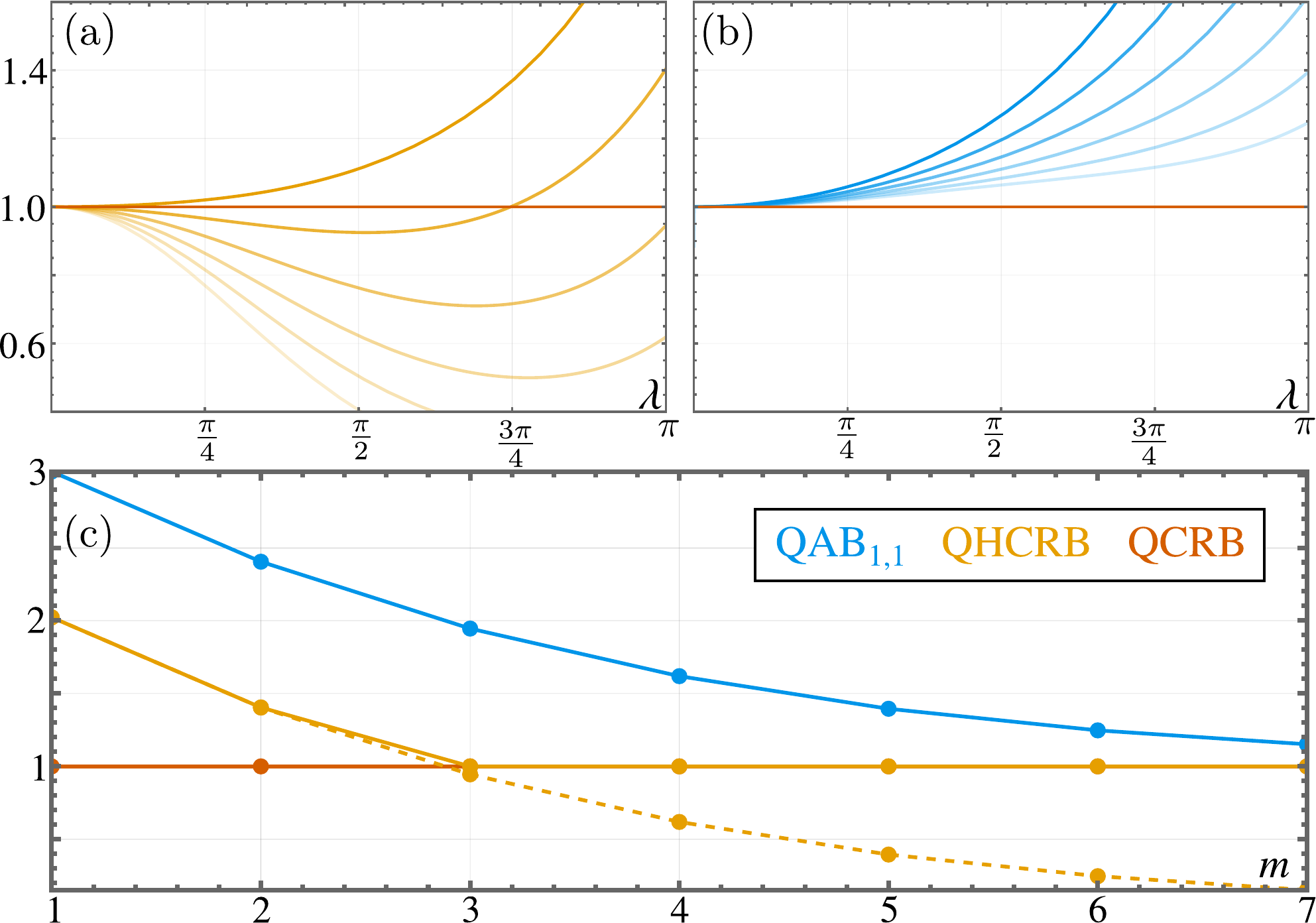}
    \caption{Bounds on the variance of an unbiased estimator in the interval $\Theta=(-\pi,\pi]$ for $m$ independent measurements on a noisy qubit with entropy $S(\rho)=0.6$. The plots show $(\Delta\theta_{\mathrm{est}})^2_{Q_{\mathrm{A}}}/(\Delta\theta_{\mathrm{est}})^2_{\mathrm{QCRB}}$ for the QABs of order $(1,1)$, $(1,0)=\mathrm{QHCRB}$ and $(0,1)=\mathrm{QCRB}$. The QAB bounds $(1,0)$ and $(1,1)$ are obtained by maximizing over the single free parameter $\lambda$ in~(\ref{eq:QAB}) and the dependence is shown  in (a) and (b), respectively. Lighter colors indicate larger values of $m=1,\dots,7$. Below the threshold value $m=3$, the maximum appears at the edge of the parameter range $\Theta$ at $\lambda_{\max}=\pi$ and the QHCRB is larger than the QCRB (c). For $m\geq 3$, the edge value (dashed orange line) is smaller than the one obtained at $\lambda_{\max}\to 0$, see also (a). In this case, the QHCRB coincides with the QCRB. The hybrid $\mathrm{QAB}_{(1,1)}$ always takes on its maximum at $\lambda_{\max}=\pi$ and converges to the QCRB from above in the limit of large $m$.}
    \label{fig:qubits}
\end{figure}

\textit{Quantum phase estimation with a qubit.---}Let us now illustrate these bounds by applying them to the relevant example of quantum phase estimation. Our goal is to estimate the phase $\theta$ of a single qubit $\rho=\frac{1}{2}(\mathbb{I}+\bm{r}^{\top}\bm{\sigma})$ with Bloch vector $\bm{r}=(0,r,0)^{\top}$ that is imprinted by the rotation $U(\theta)=e^{-i\sigma_z\theta/2}$ as $\rho(\theta)=U(\theta)\rho U(\theta)^{\dagger}$. To study threshold behavior, we analytically determine and compare the bounds $\mathrm{QCRB}=\mathrm{QAB}_{(0,1)}$, $\mathrm{QHCRB}=\mathrm{QAB}_{(1,0)}$ and the hybrid bound $\mathrm{QAB}_{(1,1)}$ for $m$ independent copies of the same qubit, $\rho(\theta)^{\otimes m}$~\cite{Supp}. The results are shown in Fig.~\ref{fig:qubits} for up to $m=7$ qubits with entropy $S(\rho)=-\mathrm{Tr}\{\rho\ln\rho\}=0.6$, corresponding to $r=|\bm{r}|\approx 0.42$. For better comparison, all bounds are normalized with respect to the QCRB, i.e., the plot shows $(\Delta\theta_{\mathrm{est}})_Q^2/(\Delta\theta_{\mathrm{est}})_{Q_{\mathrm{CRB}}}^2$ with $(\Delta\theta_{\mathrm{est}})_{Q_{\mathrm{CRB}}}^2=1/(mr^2)$. We note that the QCRB is overly optimistic for estimators that are unbiased in the range $\Theta=(\pi,\pi]$ when $m$ is small. Both bounds QHCRB and $\mathrm{QAB}_{(1,1)}$ reveal threshold behavior: They identify larger values on the lowest possible variance of an unbiased estimator, but after sufficiently many measurements, they approach the QCRB. Since the bounds at small $m$ are determined by the edge of $\Theta$, a smaller range $\Theta$ allows for smaller variances of unbiased estimators. Furthermore, the threshold behavior is more pronounced as the entropy of the qubit grows and it disappears in the limit of pure states~\cite{Supp}.

The constant value of $m(\Delta\theta_{\mathrm{est}})_{Q_{\mathrm{CRB}}}^2$ is due to the additivity of the QFI: This reflects the asymptotic limit, where each additional measurement adds as much information as the previous one. This is not the case at low data, as is shown by the more general bounds that do not satisfy additivity. As a consequence, saturation of these bounds in the $m$-shot scenario typically requires joint measurements on all $m$ copies.

\textit{General properties.---}The quantum information matrix $Q$ is closely related to the Bures metric~\cite{Huebner1992}, which stands out as the smallest among the family of metrics that contract under quantum channels~\cite{Petz1996,LesniewskiRuskai1999}. When the $G_k$ are linear functions of $\rho(\cdot)$, the quantum information function $I_{\bm{a},\bm{\lambda}}[\rho(\cdot)]=(\bm{a}^{\top}Q\bm{a})/(\bm{a}^{\top}\bm{\lambda})^2=\mathrm{Tr}\{(\bm{a}^{\top}\bm{G})\Omega_{\rho(\theta)}(\bm{a}^{\top}\bm{G})\}/(\bm{a}^{\top}\bm{\lambda})^2$ is convex in the quantum state $\rho(\cdot)$ as a consequence of the joint convexity of the Bures inner product~\cite{LiebRuskai1974,LesniewskiRuskai1999,Ruskai2007,Supp}. The bounds~(\ref{eq:qBound}) are furthermore bounded from below by the QCRB and therefore subject to the same separability limits that were derived using the QFI~\cite{PezzeSmerziReview,RenPRL2021}. We show in Ref.~\cite{Supp} that in the limit of pure states, all QABs converge to the QCRB. We further provide explicit expressions for $Q$ from the spectral decomposition of $\rho(\theta)$, from matrix vectorization techniques, and from the Bloch vector in the case of two-level systems. In the following we present an equivalent formulation of these bounds as optimization problems with constraints.

\textit{Locally best unbiased quantum parameter estimation.---}
The bounds~(\ref{eq:qBound}) are tight in the sense that they identify the minimum variance at $\theta$ by optimization over all measurements and over all estimators that satisfy the unbiasedness constraints determined by $\bm{G}$ and $\bm{\lambda}$. In other words, the bound~(\ref{eq:qBound}) is the solution to~\cite{Supp}
\begin{align}\label{eq:optimization}
    &\min_{\{E_x\}}\min_{\theta_{\mathrm{est}}} (\Delta\theta_{\mathrm{est}})^2,\notag\\
    &\mathrm{s.t.}\quad \sum_{x\in X_+}\mathrm{Tr}\{E_x\bm{G}\}(\theta_{\mathrm{est}}(x)-\langle\theta_{\mathrm{est}}\rangle_{\theta})=\bm{\lambda}.
\end{align}
Moreover, the estimator that achieves the second minimum is unique. For estimators that are unbiased throughout $\Theta$, the locally best variance is identified by the quantum Barankin bound in the limit $n\to\infty$, which demands unbiasedness at all points in $\Theta$. All other bounds discussed here are approximations of this limit. This extends Barankin's 
result~\cite{Barankin1949,Glave1972} to the quantum realm by optimization over all measurements.

\textit{Conclusions.---}We derived hierarchies of generalized quantum bounds on the variance of unbiased estimators in quantum metrology from unbiasedness constraints. The bounds converge towards the QCRB from above in two limits: When large amounts of data, i.e., many copies of the state, are available, and when the state becomes pure. For few-shot measurements with mixed states, the more general bounds reveal tighter constraints on the precision of unbiased quantum parameter estimation than the QCRB. We identify the optimal measurement observable and the estimator that achieves the smallest variance, given a set of unbiasedness constraints. Besides leading to important generalizations of the QFI and the QCRB, these bounds are useful to study threshold behavior in quantum measurements and to identify error bounds when regularity assumptions of the QCRB cannot be fulfilled.

\begin{acknowledgments}
\textit{Acknowledgments.---}We would like to thank L. Pezz\`e for fruitful discussions. This work was funded by MCIN/AEI/10.13039/501100011033 and the European Union “NextGenerationEU” PRTR fund [RYC2021-031094-I]. This work has been founded by the Ministry of Economic Affairs and Digital Transformation of the Spanish Government through the QUANTUM ENIA project call - QUANTUM SPAIN project, by the European Union through the Recovery, Transformation and Resilience Plan - NextGenerationEU within the framework of the Digital Spain 2026 Agenda, and by the CSIC Interdisciplinary Thematic Platform (PTI+) on Quantum Technologies (PTI-QTEP+). This work was supported by the European Commission through the H2020 QuantERA ERA-NET Cofund in Quantum Technologies project “MENTA”.
\end{acknowledgments}

    \clearpage

    \begin{center}\large{\textbf{Supplemental Material}}\end{center}
    \renewcommand{\theequation}{S.\arabic{equation}}
    \setcounter{equation}{0}

    \section{Derivation of the quantum bounds}
    We first derive the classical bounds~(2) and then optimize them over all POVMs to obtain the quantum bounds~(5), in the main text. 
    
    \textbf{Classical bounds for arbitrary test functions.}\\
    The starting point for this derivation is the variance of an estimator $\theta_{\mathrm{est}}$, which we write as
    \begin{align}
        (\Delta\theta_{\mathrm{est}})^2=\sum_{x\in X_+} p(x|\theta)\left(\theta_{\mathrm{est}}(x)-\langle\theta_{\mathrm{est}}\rangle_{\theta}\right)^2.
    \end{align}
    Without restriction we limit the sum to those values $x\in X_+$, with $X_+=\{x\:|\:p(x|\theta)>0\}$, that occur with nonzero probability. We now consider a family of test functions $\bm{g}(x)=(g_1(x),\dots,g_n(x))^{\top}$ that we combine into a linear combination
    \begin{align}
        \bm{a}^{\top}\bm{g}(x)=\sum_{k=1}^na_kg_k(x),
    \end{align}
    with coefficient vector $\bm{a}=(a_1,\dots,a_n)^{\top}\in\mathbb{R}^n$. From the Cauchy-Schwarz inequality
    \begin{align}\label{eq:CS}
        \left(\sum_{x} u(x)v(x)\right)^2\leq \left(\sum_{x} u(x)^2\right)\left(\sum_{x} v(x)^2\right),
    \end{align}
    with
    \begin{align}
        u(x)&=\frac{1}{\sqrt{p(x|\theta)}}\bm{a}^{\top}\bm{g}(x)\\
        v(x)&=\sqrt{p(x|\theta)}\left(\theta_{\mathrm{est}}(x)-\langle\theta_{\mathrm{est}}\rangle_{\theta}\right),
    \end{align}
    we obtain
    \begin{align}\label{eq:cBoundaApp}
        (\Delta\theta_{\mathrm{est}})^2\geq \frac{\left(\bm{a}^{\top}\bm{\lambda}\right)^2}{\bm{a}^{\top}C\bm{a}},
    \end{align}
    where $C$ is a real symmetric $n\times n$ matrix with elements
    \begin{align}\label{eq:cMatrixApp}
        C_{kl}=\sum_{x\in X_+}\frac{g_k(x)g_l(x)}{p(x|\theta)},
    \end{align}
    and
    \begin{align}\label{eq:lambdavec}
        \bm{\lambda}=\sum_{x\in X_+}\bm{g}(x)\left(\theta_{\mathrm{est}}(x)-\langle\theta_{\mathrm{est}}\rangle_{\theta}\right).
    \end{align}
    The result~(\ref{eq:cBoundaApp}) identifies a lower bound on the variance that holds for any choice of the coefficients $\bm{a}$. The tightest bound in this family is identified after maximizing over all possible choices of $\bm{a}$, i.e., $(\Delta\theta_{\mathrm{est}})^2\geq (\Delta\theta_{\mathrm{est}})_C^2$ with
    \begin{align}\label{eq:cBoundasupApp}
        (\Delta\theta_{\mathrm{est}})_C^2 = \sup_{\bm{a}} \frac{\left(\bm{a}^{\top}\bm{\lambda}\right)^2}{\bm{a}^{\top}C\bm{a}}.
    \end{align}
    Assuming $C$ to be invertible, the optimization can be carried out explicitly. We obtain from another application of the Cauchy-Schwarz inequality~(\ref{eq:CS}), this time with $\bm{u}=C^{1/2}\bm{a}$ and $\bm{v}=C^{-1/2}\bm{\lambda}$, that
    \begin{align}
        \frac{\left(\bm{a}^{\top}\bm{\lambda}\right)^2}{\bm{a}^{\top}C\bm{a}}\leq\bm{\lambda}^{\top}C^{-1}\bm{\lambda}
    \end{align}
    holds for arbitrary choices of $\bm{a}$. The maximum is achieved for the optimal $\bm{a}$ that is proportional to $C^{-1}\bm{\lambda}$:
    \begin{align}\label{eq:CSmaxaApp}
        \max_{\bm{a}}\frac{\left(\bm{a}^{\top}\bm{\lambda}\right)^2}{\bm{a}^{\top}C\bm{a}}=\bm{\lambda}^{\top}C^{-1}\bm{\lambda}.
    \end{align}
    In summary, we find that $(\Delta\theta_{\mathrm{est}})^2\geq (\Delta\theta_{\mathrm{est}})^2_C$, where
    \begin{align}\label{eq:cBoundApp}
        (\Delta\theta_{\mathrm{est}})^2_C = \bm{\lambda}^{\top}C^{-1}\bm{\lambda}.
    \end{align}
    This is the classical bound which was given in Eq.~(2) in the main text. Similar bounds for fluctuating parameters $\theta$ have been derived in Ref.~\cite{VanTreesBOOK1968}. The bound~(\ref{eq:cBoundApp}) is saturated when
    \begin{align}\label{eq:optimalitycondition}
        \bm{\lambda}^{\top}C^{-1}\bm{g}(x)&=\alpha p(x|\theta)\left(\theta_{\mathrm{est}}(x)-\langle\theta_{\mathrm{est}}\rangle_{\theta}\right)
    \end{align}
    holds with a normalization constant $\alpha\in\mathbb{R}$. The condition~(\ref{eq:lambdavec}) requires that $\alpha=1$.
    
    Choices of $g_k(x)$ that lead to a singular matrices $C$ produce divergent bounds. To avoid this, such choices are typically excluded through regularity conditions. As we will see below, these are particularly relevant when the $g_k(x)$ contain derivatives of the probability distribution.
    
    \textbf{Quantum bounds for arbitrary test observables.}\\
    For the derivation of the quantum bounds, we assume that the test functions correspond to a family of test observables $G_k$, in combination with the chosen POVM $\{E_x\}$, such that
    \begin{align}\label{eq:qgkApp}
        g_k(x)=\mathrm{Tr}\{E_xG_k\}.    
    \end{align}
    
    Furthermore, we consider the $\bm{\lambda}$ fixed, generalized conditions on the properties of the estimator $\theta_{\mathrm{est}}$, which typically address the bias. Our goal is to identify quantum bounds on the ultimate precision limits among all estimators with fixed properties $\bm{\lambda}$. The quantum bounds are given as the smallest possible lower bound~(\ref{eq:cBoundApp}) that can be achieved from an optimally chosen POVM. The following derivation generalizes the approach of Ref.~\cite{BraunsteinPRL1994} to the quantum Fisher information.
    
    We thus determine $(\Delta\theta_{\mathrm{est}})^2\geq (\Delta\theta_{\mathrm{est}})^2_Q$, with
    \begin{align}
        (\Delta\theta_{\mathrm{est}})^2_Q=\inf_{\{E_{x}\}}\bm{\lambda}^{\top}C^{-1}\bm{\lambda}.
    \end{align}
    Using~(\ref{eq:CSmaxaApp}), we obtain
    \begin{align}\label{eq:qBoundDefApp}
        (\Delta\theta_{\mathrm{est}})^2_Q=\max_{\bm{a}}\frac{\left(\bm{a}^{\top}\bm{\lambda}\right)^2}{\sup_{\{E_{x}\}}\bm{a}^{\top}C\bm{a}}.
    \end{align}
    We thus proceed to determine the maximum over all POVMs of the quantity
    \begin{align}\label{eq:aCaApp}
        \bm{a}^{\top}C\bm{a}=\sum_{x\in X_+}\frac{\left(\sum_{k=1}^na_kg_k(x)\right)^2}{p(x|\theta)}=\sum_{x\in X_+}\frac{\left(\mathrm{Tr}\{E_{x}G_{\bm{a}}\}\right)^2}{\mathrm{Tr}\{E_x\rho(\theta)\}},
    \end{align}
    where we introduced
    \begin{align}
        G_{\bm{a}}=\sum_{k=1}^na_kG_k.    
    \end{align}
    
    For an operator $\rho$ with the spectral decomposition $\rho=\sum_{j}p_j|j\rangle\langle j|$, we define the linear superoperator $\Omega_{\rho}$ through its action on an arbitrary operator $X$ as
    \begin{align}\label{eq:OmegaApp}
        \Omega_{\rho}(X)=\sum_{\substack{i,j\\p_i+p_j>0}}\frac{2}{p_i+p_j}|i\rangle\langle i|X|j\rangle\langle j|.
    \end{align}
    Note that this operator is by construction Hermitian. The special case $\Omega_{\rho(\theta)}(\frac{\partial\rho(\theta)}{\partial\theta})$ is known as the symmetric logarithmic derivative~\cite{BraunsteinPRL1994}. For the present family of bounds, we consider arbitrary operators $X$ and additional requirements need to be checked which turn out to be automatically satisfied by the symmetric logarithmic derivative. Specifically, it is straightforward to verify that the operator $\Omega_{\rho}(X)$ has the property
    \begin{align}\label{eq:OmegaPropertyGeneral}
        \frac{1}{2}\left(\Omega_{\rho}(X)\rho+\rho\Omega_{\rho}(X)\right)=X-\Pi_{\perp}X\Pi_{\perp},
    \end{align}
    where
    \begin{align}\label{eq:PiperpApp}
        \Pi_{\perp}=\sum_{\substack{j\\p_j = 0}}|j\rangle\langle j|.
    \end{align}
    When $\Pi_{\perp}X\Pi_{\perp}=0$ the operator $\Omega_{\rho}(X)$ describes to a symmetric ``division'' of $X$ by $\rho$, since we recover the original operator $X$ from $\Omega_{\rho}(X)$ by symmetrically multiplying by $\rho$. It is easy to see that the condition $\Pi_{\perp}X\Pi_{\perp}=0$ holds when $\rho=\rho(\theta)$ and $X=\frac{\partial}{\partial\theta}\rho(\theta)$, i.e., in the case of Ref.~\cite{BraunsteinPRL1994}.
    
    Using~(\ref{eq:OmegaPropertyGeneral}) in Eq.~(\ref{eq:aCaApp}), we obtain
    \begin{align}
        \bm{a}^{\top}C\bm{a}=\sum_{x\in X_+}\frac{\left(\mathrm{Tr}\{E_{x}\frac{1}{2}\left(\Omega_{\rho(\theta)}(G_{\bm{a}})\rho(\theta)+\rho(\theta)\Omega_{\rho(\theta)}(G_{\bm{a}})\right)\}\right)^2}{\mathrm{Tr}\{E_x\rho(\theta)\}},
    \end{align}
    and we assume that, for all $k$,
    \begin{align}\label{eq:suppGApp}
        \Pi_{\perp}G_k\Pi_{\perp}=0.    
    \end{align}
    The Cauchy-Schwarz inequality in the form $|\mathrm{Tr}\{A^{\dagger}B\}|^2\leq \mathrm{Tr}\{A^{\dagger}A\}\mathrm{Tr}\{B^{\dagger}B\}$ now leads to
    \begin{align}\label{eq:aCainterApp}
        \bm{a}^{\top}C\bm{a} &\stackrel{(i)}{\leq}\sum_{x\in X_+}\frac{\left|\mathrm{Tr}\{\rho(\theta)E_{x}\Omega_{\rho(\theta)}(G_{\bm{a}})\}\right|^2}{\mathrm{Tr}\{E_{x}\rho(\theta)\}}\notag\\
        &\stackrel{(ii)}{\leq}\sum_{x\in X_+}\mathrm{Tr}\{E_{x}\Omega_{\rho(\theta)}(G_{\bm{a}})\rho(\theta)\Omega_{\rho(\theta)}(G_{\bm{a}})\}\notag\\
        &\stackrel{(iii)}{\leq}\mathrm{Tr}\{\Omega_{\rho(\theta)}(G_{\bm{a}})\rho(\theta)\Omega_{\rho(\theta)}(G_{\bm{a}})\}.
    \end{align}
    The saturation conditions are
    \begin{align}\label{eq:condiApp}
        (i)\qquad\mathrm{Im}\{\mathrm{Tr}\{\rho(\theta)E_{x}\Omega_{\rho(\theta)}(G_{\bm{a}})\}\}=0,
    \end{align}
    and
    \begin{align}\label{eq:condiiApp}
        (ii)\qquad E_x^{1/2}\rho(\theta)^{1/2}=\alpha E_x^{1/2}\Omega_{\rho(\theta)}(G_{\bm{a}})\rho(\theta)^{1/2},
    \end{align}
    for some $\alpha\in\mathbb{R}$. The third inequality follows from $\sum_{x\in X_+}E_x\leq \mathbb{I}$, and is saturated when
    \begin{align}\label{eq:condiiiApp}
        (iii)\qquad \sum_{x\in X_+}E_x=\mathbb{I},
    \end{align}
    or equivalently if $p(x|\theta)>0$ for all possible values of $x$. 
    
    Using Eq.~(\ref{eq:OmegaPropertyGeneral}), we may express the result~(\ref{eq:aCainterApp}) as
    \begin{align}\label{eq:maxEaApp}
        \sup_{\{E_x\}}\bm{a}^{\top}C\bm{a}&=\mathrm{Tr}\{G_{\bm{a}}\Omega_{\rho(\theta)}(G_{\bm{a}})\},\notag\\
        &=\bm{a}^{\top}Q\bm{a},
    \end{align}
    we introduced the symmetric $n\times n$ matrix $Q$ with elements
    \begin{align}\label{eq:qMatrixApp}
        Q_{kl}=\mathrm{Tr}\{G_k\Omega_{\rho(\theta)}(G_l)\}.
    \end{align}
    
    The relation~(\ref{eq:maxEaApp}) holds for all $\bm{a}\in\mathbb{R}^n$, and inserting into Eq.~(\ref{eq:qBoundDefApp}) finally yields
    \begin{align}\label{eq:qBoundasupApp}
        (\Delta\theta_{\mathrm{est}})^2_Q=\sup_{\bm{a}}\frac{\left(\bm{a}^{\top}\bm{\lambda}\right)^2}{\bm{a}^{\top}Q\bm{a}}.
    \end{align}
    When $Q$ is invertible, we may proceed to write
    \begin{align}\label{eq:qBoundApp}
        (\Delta\theta_{\mathrm{est}})^2_Q=\bm{\lambda}^{\top}Q^{-1}\bm{\lambda}.
    \end{align}
    This is the quantum bound that was given in Eq.~(5) in the main text.
    
    The bound~(\ref{eq:qBoundApp}) can be saturated if the conditions $(i)-(iii)$ in Eqs.~(\ref{eq:condiApp})-(\ref{eq:condiiiApp}) are fulfilled for the $G_{\bm{a}}$ with optimal coefficients $\bm{a}$ that achieve the maximum in Eq.~(\ref{eq:qBoundasupApp}); in the case of $Q$ invertible, the optimal $\bm{a}$ is proportional to $Q^{-1}\bm{\lambda}$. In analogy to the QCRB~\cite{BraunsteinPRL1994}, conditions $(i)$ and $(ii)$ in Eqs.~(\ref{eq:condiApp}) and~(\ref{eq:condiiApp}) can be saturated by choosing the $E_x$ as rank-1 projectors onto the eigenstates of the Hermitian observable $\Omega_{\rho(\theta)}(G_{\bm{a}})$.
    
    Note that the saturation of the quantum bound is achieved in two steps: First, by choosing a measurement such that the classical bound coincides with the quantum bound, and, second, by choosing an estimator that reaches the classical bound. The above derivation addresses only the first step. It is in the second step that we may observe threshold behavior, i.e., difficulty to construct a practical estimator that reaches the classical bound from a small data set.
    
    \section{Locally best unbiased quantum parameter estimation}
    The bound~(\ref{eq:qBoundApp}) can also be derived from a constrained optimization problem: It coincides with the smallest variance of any estimator that satisfies a set of unbiasedness constraints. These constraints are determined by the test observables $\bm{G}$ together with the vector $\bm{\lambda}$ of constants. The solution to the optimization problem
    \begin{align}\label{eq:optimizationS}
        &\min_{\{E_x\}}\min_{\theta_{\mathrm{est}}} (\Delta\theta_{\mathrm{est}})^2,\notag\\
        &\mathrm{s.t.}\quad \sum_{x\in X_+}\mathrm{Tr}\{E_x\bm{G}\}(\theta_{\mathrm{est}}(x)-\langle\theta_{\mathrm{est}}\rangle_{\theta})=\bm{\lambda}
    \end{align}
    is given by $(\Delta\theta_{\mathrm{est}})^2_Q=\bm{\lambda}^{\top}Q^{-1}\bm{\lambda}$ and the estimator that achieves this minimum variance is unique.
    
    The ``best'' estimator, i.e., the one achieving the minimal variance for a fixed choice of POVM can be identified from the saturation condition~(\ref{eq:optimalitycondition})
    \begin{align}\label{eq:optestcl}
        \theta_{\mathrm{est}}(x)-\langle \theta_{\mathrm{est}}\rangle_{\theta}=\frac{1}{p(x|\theta)}\bm{g}(x)^{\top}C^{-1}\bm{\lambda}.
    \end{align}
    For the optimal POVM $\{E_x\}$, we obtain
    \begin{align}
        \theta_{\mathrm{est}}(x)-\langle \theta_{\mathrm{est}}\rangle_{\theta}=\frac{1}{p(x|\theta)}\bm{g}(x)^{\top}Q^{-1}\bm{\lambda},
    \end{align}
    with $p(x|\theta)=\mathrm{Tr}\{E_x\rho(\theta)\}$ and $g_k(x)=\mathrm{Tr}\{E_xG_k\}$. These expressions may be of practical relevance when the optimal estimator does not depend explicitly on $\theta$.
    
    It is straightforward to verify that this estimator is the solution that achieves the minimum in~(\ref{eq:optimizationS}). We focus on the classical case~(\ref{eq:optestcl}), since it includes the quantum-optimal one as a special case when we choose the optimal POVM and $C=Q$. First, we can easily confirm that the estimator satisfies the constraint, using that the information matrix can be expressed as
    \begin{align}
        C=\sum_{x\in X_+}\frac{\bm{g}(x)\bm{g}(x)^{\top}}{p(x|\theta)}.    
    \end{align}
    Furthermore, the above derivation of the bounds demonstrates that any other estimator $\tilde{\theta}_{\mathrm{est}}$ that also satisfies the constraints in Eq.~(\ref{eq:optimizationS}) has a variance $(\Delta\tilde{\theta}_{\mathrm{est}})^2\geq (\Delta\theta_{\mathrm{est}})^2_C=\bm{\lambda}^{\top}C^{-1}\bm{\lambda}=(\Delta\theta_{\mathrm{est}})^2$. In the last step we used that the estimator~(\ref{eq:optestcl}) achieves the minimal variance. This can be verified as follows:
    \begin{align}
        (\Delta\theta_{\mathrm{est}})^2&=\sum_{x\in X_+}p(x|\theta)\left(\theta_{\mathrm{est}}(x)-\langle \theta_{\mathrm{est}}\rangle_{\theta}\right)^2\notag\\
        &=\sum_{x\in X_+}\frac{1}{p(x|\theta)}\bm{\lambda}^{\top}C^{-1}\bm{g}(x)\bm{g}(x)^{\top}C^{-1}\bm{\lambda}\notag\\
        &=\bm{\lambda}^{\top}C^{-1}\bm{\lambda}.
    \end{align}
    
    \section{Quantum Abel bounds}
    We use our general formalism to derive the quantum Abel bounds of order $(r,s)$. The Abel bounds include, as special cases, the Barankin bounds for $s=0$ and the Bhattacharyya bounds for $r=0$, which we discuss separately in the main text. Consider the family of $1+r+s$ test observables
    \begin{align}\label{eq:GtildeApp}
        \widetilde{\bm{G}}=(\rho(\theta),\rho(\theta_1),\dots,\rho(\theta_r),\frac{\partial\rho(\theta)}{\partial\theta},\dots,\frac{\partial^s\rho(\theta)}{\partial\theta^s})^{\top},
    \end{align}
    where $\theta_k\in\Theta$ for $k=1,\dots,r$. The fixed constraints read
    \begin{align}
        \widetilde{\bm{\lambda}}&=(0,\langle\theta_{\mathrm{est}}\rangle_{\theta_1}-\langle\theta_{\mathrm{est}}\rangle_{\theta},\dots,\langle\theta_{\mathrm{est}}\rangle_{\theta_r}-\langle\theta_{\mathrm{est}}\rangle_{\theta},\notag\\&\quad\frac{\partial\langle\theta_{\mathrm{est}}\rangle_{\theta}}{\partial\theta},\dots,\frac{\partial^s\langle\theta_{\mathrm{est}}\rangle_{\theta}}{\partial\theta^s})^{\top}.
    \end{align}
    For an estimator that is unbiased in the range $\Theta$, these constraints reduce to
    \begin{align}
        \widetilde{\bm{\lambda}}&=(0,\lambda_1,\dots,\lambda_r,1,0,\dots,0)^{\top},
    \end{align}
    where we denote $\lambda_k=\theta_k-\theta$. This choice in Eqs.~(\ref{eq:cBoundasupApp}), together with an optimization over the $\lambda_1,\dots,\lambda_r$ leads us to the Abel bound of order $(r,s)$~\cite{Abel1993}
    \begin{align}
        (\Delta\theta_{\mathrm{est}})_{C_A}^2 = \sup_{\substack{\lambda_1,\dots,\lambda_r\\\bm{a}\in\mathbb{R}^{r+s+1}}} \frac{\left(\bm{a}^{\top}\widetilde{\bm{\lambda}}\right)^2}{\bm{a}^{\top}\widetilde{C}_{\mathrm{A}}\bm{a}}.
    \end{align}
    We use Eq.~(\ref{eq:qgkApp}) in~(\ref{eq:cMatrixApp}) for the test observables~(\ref{eq:GtildeApp}) to identify the $(1+r+s)\times(1+r+s)$ Abel information matrix
    \begin{align}\label{eq:cAbelApp}
        \widetilde{C}_{\mathrm{A}}=\begin{pmatrix}1 & \bm{e}^{\top} & \bm{0}^{\top}\\
        \bm{e} & C_{\mathrm{Ba}} & C_{\mathrm{H}}\\
    \bm{0} & C_{\mathrm{H}}^{\top} & C_{\mathrm{Bh}} \end{pmatrix},
    \end{align}
    where $C_{\mathrm{Ba}}$ is the $r\times r$ Barankin matrix
    \begin{align}\label{eq:CBamat}
        (C_{\mathrm{Ba}})_{kl}=\sum_{x\in X_+}p(x|\theta)L(x|\theta+\lambda_k,\theta)L(x|\theta+\lambda_l,\theta),
    \end{align}
    with the likelihood ratio $L(x|\theta+\lambda_k,\theta)=\frac{p(x|\theta+\lambda_k)}{p(x|\theta)}$, $C_{\mathrm{Bh}}$ is the $s\times s$ Bhattacharyya matrix with elements
    \begin{align}
        (C_{\mathrm{Bh}})_{kl}=\sum_{x\in X_+}\frac{1}{p(x|\theta)}\left(\frac{\partial^kp(x|\theta)}{\partial\theta^k}\right)\left(\frac{\partial^lp(x|\theta)}{\partial\theta^l}\right),
    \end{align}
    and $C_{\mathrm{H}}$ is a $r\times s$ hybrid matrix with elements
    \begin{align}\label{eq:CHmat}
        (C_{\mathrm{H}})_{kl}=\sum_{x\in X_+}\frac{1}{p(x|\theta)}p(x|\theta+\lambda_k)\left(\frac{\partial^lp(x|\theta)}{\partial\theta^l}\right).
    \end{align}
    Moreover, the values that appear in the first row and column in Eq.~(\ref{eq:cAbelApp}) are easily explained: Whenever one of the $\lambda_k$ is zero, the Barankin term~(\ref{eq:CBamat}) yields $1$ and the hybrid terms~(\ref{eq:CHmat}) vanish. We express this with the vector $\bm{e}=(1,\dots,1)^{\top}\in\mathbb{R}^r$ and the zero vector $\bm{0}\in\mathbb{R}^s$. Assuming $\widetilde{C}_{\mathrm{A}}$ to be invertible, we obtain the sharpest formulation of the Abel bound from Eq.~(\ref{eq:cBoundApp}) after an optimization over the test points $\lambda_1,\dots,\lambda_r$:
    \begin{align}\label{eq:cAbelTildeApp}
        (\Delta\theta_{\mathrm{est}})^2_{C_A} = \sup_{\lambda_1,\dots,\lambda_r}\widetilde{\bm{\lambda}}^{\top}(\widetilde{C}_{\mathrm{A}})^{-1}\widetilde{\bm{\lambda}}.
    \end{align}
    We rewrite the matrix~(\ref{eq:cAbelApp}) as
    \begin{align}
        \widetilde{C}_{\mathrm{A}}=\begin{pmatrix}1 & \bm{f}^{\top} \\
            \bm{f} & C_{\mathrm{A}} \end{pmatrix},
    \end{align}
    where
    \begin{align}\label{eq:cAbelcompactApp}
        C_{\mathrm{A}}=\begin{pmatrix}C_{\mathrm{Ba}} & C_{\mathrm{H}}\\
        C_{\mathrm{H}}^{\top} & C_{\mathrm{Bh}} \end{pmatrix},
    \end{align}
    and we have combined $\bm{e}$ and $\bm{0}$ into the vector $\bm{f}=\bm{e}\oplus\bm{0}$. Using block inversion and
    $\widetilde{\bm{\lambda}}=1\oplus\bm{\lambda}$ with $\bm{\lambda}=(\lambda_1,\dots,\lambda_r,1,0,\dots,0)^{\top}\in\mathbb{R}^{r+s}$, we can rewrite the Abel bound~(\ref{eq:cAbelTildeApp}) as
    \begin{align}\label{eq:cAbelfullApp}
        (\Delta\theta_{\mathrm{est}})^2_{C_A} = \sup_{\lambda_1,\dots,\lambda_r}\bm{\lambda}^{\top}(C_{\mathrm{A}}-\bm{f}\bm{f}^{\top})^{-1}\bm{\lambda}.
    \end{align}
    Note that an explicit but lengthy expression that depends only on the smaller matrices inside the block matrix Eq.~(\ref{eq:cAbelcompactApp}) can be derived by using further block inversion techniques, see, e.g., Ref.~\cite{Abel1993}.
    
    The quantum Abel bounds are obtained from Eq.~(\ref{eq:qBoundasupApp}) and read
    \begin{align}\label{eq:QABasupApp}
        (\Delta\theta_{\mathrm{est}})_{Q_{\mathrm{A}}}^2&=\inf_{\{E_x\}}(\Delta\theta_{\mathrm{est}})_{C_{\mathrm{A}}}^2\notag\\&=
        \sup_{\substack{\lambda_1,\dots,\lambda_r\\\bm{a}\in\mathbb{R}^{r+s+1}}} \frac{\left(\bm{a}^{\top}\widetilde{\bm{\lambda}}\right)^2}{\bm{a}^{\top}\widetilde{Q}_{\mathrm{A}}\bm{a}},
    \end{align}
    where the quantum Abel information (QAI)
    \begin{align}\label{eq:qAbelApp}
        \widetilde{Q}_{\mathrm{A}}=\begin{pmatrix}1 & \bm{e}^{\top} & \bm{0}^{\top}\\
        \bm{e} & Q_{\mathrm{Ba}} & Q_{\mathrm{H}}\\
    \bm{0} & Q_{\mathrm{H}}^{\top} & Q_{\mathrm{Bh}} \end{pmatrix},
    \end{align}
    contains the $r\times r$ quantum Barankin matrix
    \begin{align}\label{eq:qBaMatrixApp}
        (Q_{\mathrm{Ba}})_{kl}=\mathrm{Tr}\{\rho(\theta+\lambda_k)\Omega_{\rho(\theta)}(\rho(\theta+\lambda_l))\},
    \end{align}
    the $s\times s$ quantum Bhattacharyya matrix
    \begin{align}\label{eq:qBhMatrixApp}
        (Q_{\mathrm{Bh}})_{kl}=\mathrm{Tr}\{\left(\frac{\partial^k\rho(\theta)}{\partial\theta^k}\right)\Omega_{\rho(\theta)}\left(\frac{\partial^l\rho(\theta)}{\partial\theta^l}\right)\},
    \end{align}
    and the $r\times s$ hybrid matrix
    \begin{align}
        (Q_{\mathrm{H}})_{kl}=\mathrm{Tr}\{\rho(\theta+\lambda_k)\Omega_{\rho(\theta)}\left(\frac{\partial^l\rho(\theta)}{\partial\theta^l}\right)\}.
    \end{align}
    To determine the first row and column, we make use of the fact
    \begin{align}
        \Omega_{\rho}(\rho)=\mathbb{I}-\Pi_{\perp},
    \end{align}
    which follows directly from the definitions~(\ref{eq:OmegaApp}) and~(\ref{eq:PiperpApp}). Furthermore, using~(\ref{eq:suppGApp}), we obtain for all $k$:
    \begin{align}
        \mathrm{Tr}\{G_k\Omega_{\rho(\theta)}(\rho(\theta))\}=\mathrm{Tr}\{\rho(\theta)\Omega_{\rho(\theta)}(G_k)\}=\sum_{\substack{j\\p_j\neq 0}}\langle j|G_k|j\rangle=\mathrm{Tr}\{G_k\}.
    \end{align}
    For the family $\widetilde{\bm{G}}$ in~(\ref{eq:GtildeApp}), this yields the first row and column of $\widetilde{Q}_{\mathrm{A}}$ as given in Eq.~(\ref{eq:qAbelApp}).
    
    In analogy to the classical case, if the matrix $\widetilde{Q}_{\mathrm{A}}$ is invertible, we may use the sharper bound
    \begin{align}\label{eq:qAbelfullApp}
        (\Delta\theta_{\mathrm{est}})_{Q_{\mathrm{A}}}^2=\sup_{\lambda_1,\dots,\lambda_r}\bm{\lambda}^{\top}(Q_{\mathrm{A}}-\bm{f}\bm{f}^{\top})^{-1}\bm{\lambda},
    \end{align}
    with 
    \begin{align}\label{eq:qABMatrixApp}
        Q_{\mathrm{A}}=\begin{pmatrix}Q_{\mathrm{Ba}} & Q_{\mathrm{H}}\\Q^{\top}_{\mathrm{H}}& Q_{\mathrm{Bh}} \end{pmatrix}.
    \end{align}
    
    Note that the set of test observables $\widetilde{\bm{G}}$ in Eq.~(\ref{eq:GtildeApp}) contains $\rho(\theta)$ in addition to the observables $\bm{G}$ described in the main text. This observable adds the trivial constraint $\lambda_0=0$, i.e., it does not increase the number of optimization variables for the Barankin terms. However, it leads to the subtraction of the term $\bm{f}\bm{f}^{\top}$ in Eqs.~(\ref{eq:cAbelfullApp}) and~(\ref{eq:qAbelfullApp}). In the case of the Bhattacharyya bounds, where $\bm{f}=\bm{0}$, this leaves the bound unchanged, whereas in all other cases, it leads to a tighter bound at no additional cost, since $\bm{f}\bm{f}^{\top}\geq 0$ is positive semi-definite. Hence, we implicitly always add $\rho(\theta)$ to the set of test observables.
    
    \textbf{Quantum Barankin bounds}\\
    The quantum Barankin bounds of order $r$ are contained as the special cases $s=0$ of the quantum Abel bounds $(r,s)$. Note that in this case $\bm{f}=\bm{e}$ and $\bm{\lambda}=(\lambda_1,\dots,\lambda_r)^{\top}$. The expression~(\ref{eq:qAbelfullApp}) in this case leads to the following expression for the quantum Barankin bound:
    \begin{align}\label{eq:qBafullApp}
        (\Delta\theta_{\mathrm{est}})_{Q_{\mathrm{Ba}}}^2=\sup_{\bm{\lambda}\in\Theta^r}\bm{\lambda}^{\top}(Q_{\mathrm{Ba}}-\bm{e}\bm{e}^{\top})^{-1}\bm{\lambda}.
    \end{align}
    
    \textbf{Quantum Bhattacharyya bounds}\\
    The quantum Bhattacharyya bounds of order $s$ are obtained from the quantum Abel bounds $(r,s)$ for $r=0$. In this case, $\bm{f}=\bm{0}$ and $\bm{\lambda}=(1,0,\dots,0)^{\top}$ is the first unit vector in $\mathbb{R}^s$. We obtain from~(\ref{eq:qAbelfullApp}) the quantum Bhattacharyya bound as:
    \begin{align}\label{eq:qBhfullApp}
        (\Delta\theta_{\mathrm{est}})_{Q_{\mathrm{Bh}}}^2=(Q_{\mathrm{Bh}}^{-1})_{11}.
    \end{align}
    
    We remark that the test observables $\frac{\partial^k\rho(\theta)}{\partial\theta^k}$ associated to the Bhattacharyya bounds may be undefined or lead to vanishing test functions $g_k(x)$. In this case, the classical bound can diverge. This reflects the impossibility to find an optimal estimator with a constraint that cannot be satisfied. The test observables $\rho(\theta_k)$ associated with the Barankin bounds, however, lead to nonzero $g_k(x)$ owing to the normalization of the quantum states. Hence, the Barankin bounds are able to circumvent the regularity conditions that are needed to avoid these divergences in the case of the Bhattacharyya family of bounds.

    \section{Properties and explicit expressions}
    In this section we provide tools for the explicit calculation of the quantum bounds. We further apply these tools to particular cases of interest, such as a single qubit or a pure state mixed with white noise. We also show that all QAB bounds collapse to the QCRB for pure states.
    
    \textbf{Relation to contractive Riemannian metrics.}\\
    The superoperator $\Omega_{\rho}$ is often expressed for nonsingular $\rho$ as $\Omega_{\rho}=2(L_{\rho}+R_{\rho})^{-1}$, where $L_A(X)=AX$ and $R_A(X)=XA$ are superoperators defining left- and right multiplication by $A$. It defines an inner product $\langle A,B\rangle = \mathrm{Tr}\{A^{\dagger}\Omega_{\rho}(B)\}$ and a Riemannian metric on the space of density matrices that is contractive under completely positive and trace preserving operations~\cite{Petz1996,LesniewskiRuskai1999}. It was shown that among the family of metrics with this property, the one associated with $\Omega_{\rho}$, also known as the Bures metric~\cite{Huebner1992}, is always the smallest one~\cite{LesniewskiRuskai1999}.

    The inverse of~(5) defines a quantum information function that depends via the matrix $Q$ on the quantum state $\rho(\theta)$ and the test observables $\bm{G}$, which in the cases mentioned above are again functions of $\rho(\theta_k)$ for $\theta_k\in\Theta$. For a fixed choice of the constants $\bm{\lambda}$ and $\bm{a}$, we thus obtain a quantum information function
    \begin{align}\label{eq:QIF}
        I_{\bm{a},\bm{\lambda}}[\rho(\cdot)]=\frac{\bm{a}^{\top}Q\bm{a}}{(\bm{a}^{\top}\bm{\lambda})^2},    
    \end{align}
    which is a property of the one-parameter family of quantum states $\rho(\theta)$ for $\theta\in\Theta$. 
    
    As mentioned above, the family of quantum information functions~(\ref{eq:QIF}) includes the QFI as a special case, which satisfies the important property of being a convex function of the quantum state. Interestingly, all quantum information function $I_{\bm{a},\bm{\lambda}}[\rho(\cdot)]$ can be shown to be convex in $\rho(\cdot)$ when the test observables $G_k$ are linear functions of $\rho(\cdot)$. This includes all cases that have been discussed above. Convexity follows from the joint convexity of the inner product $M_{\rho}(A,B)=\mathrm{Tr}\{A^{\dagger}\Omega_{\rho}(B)\}$ in $A,B$, and $\rho$~\cite{LiebRuskai1974,LesniewskiRuskai1999,Ruskai2007}. For $\rho=\sum_{\gamma}p_{\gamma}\rho_{\gamma}$, with non-singular density matrices $\rho_{\gamma}$, we have from linearity $G_k=\sum_{\gamma}p_{\gamma}G_{k,\gamma}$, and joint convexity implies $I_{\bm{a},\bm{\lambda}}[\sum_{\gamma}p_{\gamma}\rho_{\gamma}(\cdot)]\leq\sum_{\gamma}p_{\gamma}I_{\bm{a},\bm{\lambda}}[\rho_{\gamma}(\cdot)]$, i.e., the convexity of $I_{\bm{a},\bm{\lambda}}$.

    \textbf{From spectral decomposition.}\\
    The quantum information matrix $Q$ in Eq.~(\ref{eq:qMatrixApp}) that determines the quantum bounds~(5) and the optimal measurement can be explicitly determined from the spectral decomposition of $\rho(\theta)=\sum_ip_i|i\rangle\langle i|$. Using Eq.~(\ref{eq:OmegaApp}) in Eq.~(\ref{eq:qMatrixApp}) yields
    \begin{align}\label{eq:QspectralApp}
        Q_{kl}=\sum_{\substack{i,j\\p_i+p_j>0}}\frac{2}{p_i+p_j}\langle i|G_k|j\rangle\langle j|G_l|i\rangle.
    \end{align}
    
    \textbf{From matrix vectorization.}\\
    We may use matrix vectorization techniques to write the Hilbert-Schmidt product between two operators as $\mathrm{Tr}\{A^{\dagger}B\}=\langle\langle A|B\rangle\rangle$, where $|A\rangle\rangle=\sum_{ij}a_{ij}|i\rangle\otimes|j\rangle$ is the vectorization of the operator $A=\sum_{ij}a_{ij}|i\rangle\langle j|$. Using the relation $|AOB\rangle\rangle = (B^{\top}\otimes A)|O\rangle\rangle$ yields the representation of $\Omega_{\rho}$ on the vector space $\tilde{\Omega}_{\rho}=\frac{2}{\rho(\theta)\otimes\mathbb{I}+\mathbb{I}\otimes\rho(\theta)^{\top}}$~\cite{TemmeJMP2015} and the expression
    \begin{align}
        Q_{kl}=\langle\langle G_k|\frac{2}{\rho(\theta)\otimes\mathbb{I}+\mathbb{I}\otimes\rho(\theta)^{\top}}|G_l\rangle\rangle.
    \end{align}
    In some situations, matrix vectorization may be more convenient than diagonalization of $\rho(\theta)$, even though it comes at the expense of a Hilbert space of larger dimension. 
    
    \textbf{From the Bloch vector.}\\
    For a nonsingular single-qubit state, we can express $Q_{kl}$ explicitly in terms of the Bloch vector $\bm{r}_{\theta}$ of the state $\rho(\theta)=\frac{1}{2}(\mathbb{I}+\bm{r}_{\theta}^{\top}\bm{\sigma})$, where $\bm{\sigma}=(\sigma_x,\sigma_y,\sigma_z)^{\top}$ is a vector of Pauli matrices. We obtain
    \begin{align}\label{eq:Qqubit}
        Q_{kl}=\frac{(g_{k}^0-\bm{g}_k^{\top}\bm{r}_{\theta})(g_{l}^0-\bm{g}_l^{\top}\bm{r}_{\theta})}{1-|\bm{r}_{\theta}|^2}+\bm{g}_k^{\top}\bm{g}_l,
    \end{align}
    where $G_k=\frac{1}{2}(g_{k}^0\mathbb{I}+\bm{g}_{k}^{\top}\bm{\sigma})$.
    
    The elements of the quantum Barankin matrix are obtained from the test observables $G_k=\frac{1}{2}(\mathbb{I}+\bm{r}_{\theta_k}^{\top}\bm{\sigma})$, the quantum Bhattacharyya matrix elements correspond to $G_k=\frac{1}{2}(\frac{\partial^k\bm{r}_{\theta}}{\partial \theta^k})^{\top}\bm{\sigma}$, and the hybrid matrix $Q_{\mathrm{H}}$ contains the crossterms that involve both types.

    \begin{figure}[tb]
        \centering
        \includegraphics[width=.48\textwidth]{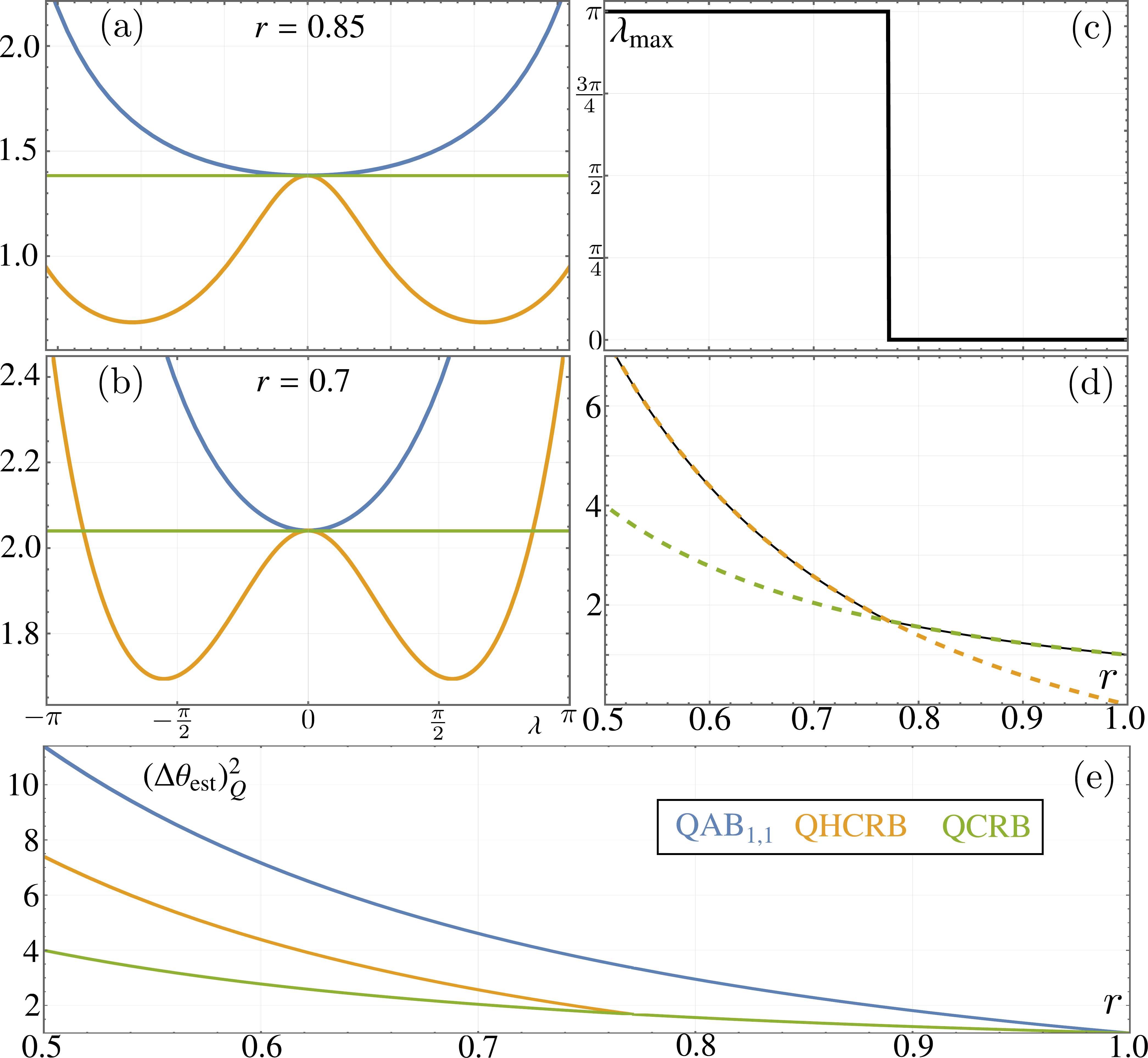}
        \caption{Single-shot bounds on the variance of an unbiased estimator in the interval $\Theta=(-\pi,\pi]$ for noisy qubits with Bloch vector length $r=|\vec{r}|$. We compare the QABs of order $(1,0)=\mathrm{QHCRB}$ and $(0,1)=\mathrm{QCRB}$ with the hybrid bound of order $(1,1)$. The bounds are obtained by maximizing over the single free parameter $\lambda$ in~(15) and the dependence is shown for $r=0.85$ and $r=0.7$, in panels (a) and (b), respectively. For states with high purity, the QHCRB is maximized by $\lambda_{\max}\to 0$ and, in this case, coincides with the QCRB. Below a threshold purity, a new maximum emerges at the extreme of the parameter range $\Theta$ at $\lambda=\pi$ (c) and the QHCRB is larger than the QCRB (d). The hybrid $\mathrm{QAB}_{(1,1)}$ provides the largest lower bound on the variance of an unbiased estimator, while all bounds coincide in the limit of pure states, $r\to 1$ (e).}
        \label{fig:qubitSupp}
    \end{figure}
    
    \section{Phase estimation with a single qubit}
    To illustrate applicability of the bounds, we consider the paradigmatic case of phase estimation by repeated measurements of a two-level system. The phase parameter is imprinted by the unitary evolution $U_{\bm{n}}(\theta)=\exp(-i\bm{n}^{\top}\bm{\sigma}\theta/2)$. The rotated qubit state $\rho(\theta)=U_{\bm{n}}(\theta)\rho(0)U^{\dagger}_{\bm{n}}(\theta)=\frac{1}{2}(\mathbb{I}+\bm{r}_{\theta}^{\top}\bm{\sigma})$ can be described in terms of the evolution of its Bloch vector:
    \begin{align}\label{eq:BVevo}
        \bm{r}_{\theta}=(\cos\theta)(\bm{r}_0-(\bm{n}^{\top}\bm{r}_0)\bm{n})+(\bm{n}^{\top}\bm{r}_0)\bm{n}+(\sin\theta)\bm{n}\times\bm{r}_0,
    \end{align}
    where $\bm{r}_0$ is the Bloch vector of the initial state $\rho_{0}$.
    
    For the single-shot scenario, $m=1$, we may combine Eq.~(\ref{eq:BVevo}) with Eq.~(\ref{eq:Qqubit}) to determine the QAB of any order. Here, we limit ourselves to the QABs of order $(0,1)$, $(1,0)$, and $(1,1)$, which correspond respectively to the QCRB (the first-order QBhB), the QHCRB (the first-order QBaB) and a hybrid bound containing the first order of both approaches. For the case discussed in the main text, i.e., $\vec{r}=(0,r,0)^{\top}$ and $\vec{n}=(0,0,1)^{\top}$, Fig.~\ref{fig:qubitSupp} shows how these bounds depend on the entropy $S(\rho)$ of a single qubit.
    
    For $m$-shot scenarios, we replace $\rho(\theta)$ by $\rho(\theta)^{\otimes m}$ and use Eq.~(\ref{eq:QspectralApp}) based on the spectral decomposition of $\rho(\theta)^{\otimes m}$. The single qubit decomposes as
    \begin{align}
        \rho(\theta)=p_+|+\rangle\langle +|+p_-|-\rangle\langle -|,
    \end{align}
    with $p_{\pm}=\frac{1\pm |\bm{r}_{\theta}|}{2}$ and $|\pm\rangle\langle\pm|=\frac{1}{2}(\mathbb{I}\pm \frac{\bm{r}_{\theta}^{\top}}{|\bm{r}_{\theta}|}\bm{\sigma})$. Using
    \begin{align}
        \Omega_{\rho(\theta)}(X)&=\sum_{\bm{\epsilon},\bm{\eta}}\frac{2}{p_{\bm{\epsilon}}+p_{\bm{\eta}}}|\bm{\epsilon}\rangle\langle \bm{\epsilon}|X|\bm{\eta}\rangle\langle \bm{\eta}|,\notag
    \end{align}
    with $\bm{\epsilon}=(\epsilon_1,\dots,\epsilon_m)$, $p_{\bm{\epsilon}}=p_{\epsilon_1}\cdots p_{\epsilon_m}$ and $|\bm{\epsilon}\rangle=|\epsilon_1\rangle\otimes\cdots\otimes|\epsilon_m\rangle$, and $\epsilon_i=\pm$ for $i=1,\dots,m$, we determine the elements of the QAB for $m$ shots. The elements of the Barankin matrix read
    \begin{widetext}
    \begin{align}
        (Q_{\mathrm{Ba}}^{(m)})_{kl}&=\sum_{\bm{\epsilon},\bm{\eta}}\frac{2}{p_{\bm{\epsilon}}+p_{\bm{\eta}}}\prod_{s=1}^m\frac{1}{2^3}\left[1+\epsilon_s\eta_s+(\epsilon_s+\eta_s)\frac{\bm{r}_{\theta}}{|\bm{r}_{\theta}|}\cdot(\bm{r}_{\theta+\lambda_k}+\bm{r}_{\theta+\lambda_l})+(1-\epsilon_s\eta_s)\bm{r}_{\theta+\lambda_k}\cdot\bm{r}_{\theta+\lambda_l}\right.\notag\\&\hspace{6cm}\left.+i\frac{\epsilon_s-\eta_s}{|\bm{r}_{\theta}}(\bm{r}_{\theta}\times\bm{r}_{\theta+\lambda_l})\cdot\bm{r}_{\theta+\lambda_k}+2\frac{\epsilon_s\eta_s}{|\bm{r}_{\theta}|^2}(\bm{r}_{\theta}\cdot\bm{r}_{\theta+\lambda_k})(\bm{r}_{\theta}\cdot\bm{r}_{\theta+\lambda_l})\right].
    \end{align}
    \end{widetext}
    The first-order element of the Bhattacharyya matrix coincides with the QFI, which is additive. We thus obtain $(Q_{\mathrm{Bh}}^{(m)})_{11}=m(Q_{\mathrm{Bh}}^{(1)})_{11}$, which can be easily determined using Eq.~(\ref{eq:Qqubit}). Finally, the first-order hybrid term can also shown to be additive:
    \begin{align}
        (Q_{\mathrm{H}}^{(m)})_{k1}=\sum_{\bm{\epsilon},\bm{\eta}}\frac{2}{p_{\bm{\epsilon}}+p_{\bm{\eta}}}\langle \bm{\epsilon}|\rho(\theta+\lambda_k)^{\otimes m}|\bm{\eta}\rangle\langle\bm{\eta}|\frac{\partial\rho(\theta)^{\otimes m}}{\partial\theta}|\bm{\epsilon}\rangle.
    \end{align}
    Using
    \begin{align}
        \langle\bm{\eta}|\frac{\partial\rho(\theta)^{\otimes m}}{\partial\theta}|\bm{\epsilon}\rangle&=\sum_{i=1}^m\langle\epsilon_i|\frac{\partial\rho(\theta)}{\partial\theta}|\eta_i\rangle\prod_{\substack{j=1\\(j\neq i)}}^mp_{\epsilon_j}\delta_{\epsilon_j\eta_j},
    \end{align}
    and $\sum_{\epsilon_j}\langle\epsilon_j|\rho(\theta+\lambda_k)|\epsilon_j\rangle=1$, we obtain
    \begin{align}
        (Q_{\mathrm{H}}^{(m)})_{k1}&=m\sum_{\epsilon,\eta=\pm}\frac{2}{p_{\epsilon}+p_{\eta}}\langle\epsilon|\rho(\theta+\lambda_k)|\eta\rangle\langle\eta|\frac{\partial\rho(\theta)}{\partial\theta}|\epsilon\rangle\notag\\
        &=m(Q_{\mathrm{H}}^{(1)})_{k1},
    \end{align}
    which can again be determined from Eq.~(\ref{eq:Qqubit}).
    
    \section{Limit of pure states}
    The quantum bounds apply to test observables $G_k$ that satisfy the condition~(\ref{eq:suppGApp}). Since the operator $\Omega_{\rho(\theta)}$ represents a ``division'' by $\rho(\theta)$, singularities may arise when some eigenvalues vanish. To see the emergence of these singularities, we consider the case of a pure state mixed with white noise
    \begin{align}\label{eq:whitenoisyrho}
    \rho(\theta)=(1-\epsilon)|\Psi(\theta)\rangle\langle\Psi(\theta)|+\frac{\epsilon}{d}\mathbb{I}
    \end{align}
    in the definition~(\ref{eq:OmegaApp}), where $d$ is the Hilbert space dimension. The density matrix~(\ref{eq:whitenoisyrho}) has the nondegenerate eigenvalue $1+\epsilon(d^{-1}-1)$ with eigenvector $|\Psi(\theta)\rangle$ and the $(d-1)$-fold degenerate eigenvalue $\epsilon d^{-1}$ with eigenvectors that span the space orthogonal to $|\Psi(\theta)\rangle$. We obtain from Eq.~(\ref{eq:QspectralApp}):
    \begin{align}\label{eq:OmegaEpsilonApp}
        \Omega_{\rho(\theta)}(X)
        &=\left(\frac{1}{1-\epsilon+\frac{\epsilon}{d}}-\frac{4}{1-\epsilon+2\frac{\epsilon}{d}}\right)\Pi_{\Psi(\theta)}X\Pi_{\Psi(\theta)}\notag\\&\quad+\left(\frac{2}{1-\epsilon+2\frac{\epsilon}{d}}\right)\left(\Pi_{\Psi(\theta)}X+X\Pi_{\Psi(\theta)}\right)\notag\\&\quad+\frac{d}{\epsilon}(\mathbb{I}-\Pi_{\Psi(\theta)})X(\mathbb{I}-\Pi_{\Psi(\theta)}),
    \end{align}
    where $\Pi_{\Psi(\theta)}=|\Psi(\theta)\rangle\langle\Psi(\theta)|$ is the projector onto the pure state $|\Psi(\theta)\rangle$. The operator~(\ref{eq:OmegaEpsilonApp}) diverges in the pure-state limit $\epsilon\to 0$ due to the last term, which carries the factor $d\epsilon^{-1}$. The divergence can be avoided if this term vanishes, which happens precisely when
    \begin{align}\label{eq:XperpPureApp}
        \Pi_{\perp}X\Pi_{\perp}=0,
    \end{align}
    where $\Pi_{\perp}=\mathbb{I}-\Pi_{\Psi(\theta)}$. This is consistent with the condition~(\ref{eq:suppGApp}). In the present case of a pure state, Hermitian operators $X$ that satisfy~(\ref{eq:XperpPureApp}) have the form $X=|\Psi(\theta)\rangle\langle\varphi|+|\varphi\rangle\langle\Psi(\theta)|$ with $|\varphi\rangle$ arbitrary. The special case $X=\frac{\partial}{\partial\theta}|\Psi(\theta)\rangle\langle\Psi(\theta)|$ with $|\varphi\rangle=|\frac{\partial}{\partial{\theta}}\Psi(\theta)\rangle$ leads to the symmetric logarithmic derivative of a pure state~\cite{BraunsteinPRL1994,PezzeSmerziReview}. 
    
    What does this imply for the QAB in the limit of pure states? To answer this question, we consider the form given in Eq.~(\ref{eq:QABasupApp}). First notice that, by construction, the optimization on the right-hand-side of Eq.~(\ref{eq:QABasupApp}) cannot yield a value lower than the QCRB, since this bound can always be recovered for special choices of the $\bm{a}$ and $\lambda_1,\dots,\lambda_r$. For QABs of order ($r,s$) with $s>0$, for instance, the QFI is contained in the Bhattacharyya matrix as the first element on the diagonal, and choosing $\bm{a}$ such that all other coefficients are zero yields the QCRB. For $s=0$ and $r>0$, the QCRB can be recovered from the first-order QBaB (the QHCRB) in the limit $\lambda\to 0$. Hence, we have that $(\Delta\theta_{\mathrm{est}})^2_{Q_{\mathrm{A}}}\geq (\Delta\theta_{\mathrm{est}})^2_{Q_{\mathrm{CR}}}$, i.e., the QABs are always bounded from below by the QCRB.
    
    Second, we notice that the condition~(\ref{eq:XperpPureApp}) is not fulfilled in the case of a pure state (or for the noisy pure state~(\ref{eq:whitenoisyrho}) in the limit of $\epsilon\to 0$) for the terms $X=\rho(\theta+\lambda_k)$ that appear in the quantum Barankin matrix~(\ref{eq:qBaMatrixApp}). Test observables of the kind $\rho(\theta+\lambda_k)=|\Psi(\theta+\lambda_k)\rangle\langle\Psi(\theta+\lambda_k)|$ satisfy the condition~(\ref{eq:XperpPureApp}) if and only if $|\Psi(\theta+\lambda_k)\rangle$ coincides with $|\Psi(\theta)\rangle$ up to a global phase. Recall, however, that in the limit $\lambda_k\to 0$, the difference between two such terms becomes proportional to the derivative $\frac{\partial}{\partial\theta}|\Psi(\theta)\rangle\langle\Psi(\theta)|$, which satisfies the condition~(\ref{eq:XperpPureApp}).
    
    Similarly, also the higher-order Bhattacharyya terms violate the condition~(\ref{eq:XperpPureApp}). While for test observables $\frac{\partial^k}{\partial\theta^k}\rho(\theta)$ with derivatives of order $k=0$ or $k=1$, i.e., the two terms that are also present in the QCRB, the condition~(\ref{eq:XperpPureApp}) is satisfied, this is no longer the case for derivatives of order $k\geq 2$. The same applies to further possible extensions such as derivatives at different test points $\frac{\partial^k}{\partial\theta^k}\rho(\theta+\lambda_l)$.
    
    To conclude this argument, we find that $\Omega_{\rho(\theta)}(G_k)$ diverges in the limit of pure states for all generalized test observables $G_k$ that appear in the QAB except for $G_k=\frac{\partial}{\partial\theta}|\Psi(\theta)\rangle\langle\Psi(\theta)|$, which can always be constructed, either directly from the first-order term in the Bhattacharyya matrix if $s>0$ or from zero and first-order terms of the Barankin matrix when $r>0$, by taking the limit $\lambda\to 0$. Since all other terms grow to infinity as we approach the pure-state limit $\epsilon\to 0$ in Eq.~(\ref{eq:whitenoisyrho}), the optimization over coefficients in Eq.~(\ref{eq:QABasupApp}) converges to the QCRB, and we have
    \begin{align}
        \lim_{\epsilon\to 0}(\Delta\theta_{\mathrm{est}})_{Q_{\mathrm{A}}}^2=(\Delta\theta_{\mathrm{est}})_{Q_{\mathrm{CRB}}}^2,
    \end{align}
    with $(\Delta\theta_{\mathrm{est}})_{Q_{\mathrm{CRB}}}^2=F_Q[|\Psi(\theta)\rangle\langle\Psi(\theta)|]^{-1}$.

\end{document}